\documentclass[twocolumn]{emulateapj}

\newcommand{\Dtsamp}{{\Delta}T_{{\rm samp}}}

\newcommand{\xsqdist}{\chi ^2 _{{\rm dist}}}

\newcommand{\mbh}{M_{\rm BH}}
\newcommand{\Tb}{T_{\rm b}}

\newcommand{\xte}{{\it RXTE}}

\newcommand{\Msun}{\hbox{$\rm\thinspace M_{\odot}$}}

\slugcomment{}
\shorttitle{X-ray Power Spectrum of NGC 7469}
\shortauthors{Markowitz}
\begin{document}
\title{The X-ray Power Spectral Density Function of the Seyfert Active Galactic Nucleus NGC 7469}
\author{A.\ Markowitz}
\affil{Center for Astrophysics and Space Sciences, University of California, San Diego, Mail Code 0424, La Jolla, California, 92093-0424, USA}

\begin{abstract}
We present the broadband X-ray power spectral density function (PSD)
of the X-ray-luminous Seyfert 1.2 NGC 7469, measured from {\it Rossi
  X-ray Timing Explorer} monitoring data and two {\it XMM-Newton}
observations.  We find significant evidence for a turnover in the
2--10 keV PSD at a temporal frequency of $2.0^{+3.0}_{-0.8} \times
10^{-6}$ Hz or $1.0^{+3.0}_{-0.6} \times 10^{-6}$ Hz, depending on the
exact form of the break (sharply-broken or slowly-bending power-law,
respectively).  The ``surrogate'' Monte Carlo method of Press et
al.\ (1992) was used to map out the probability distributions of PSD
model parameters and obtain reliable uncertainties (68$\%$ confidence
limits quoted here).  The corresponding break time scale of $5.8 \pm
3.5$ days or $11.6^{+17.5}_{-8.7}$ days, respectively, is consistent
with the empirical relation between PSD break time scale, black hole
mass and bolometric luminosity of M$^{\rm c}$Hardy et al.  Compared to
the 2--10 keV PSD, the 10--20 keV PSD has a much flatter shape at high
temporal frequencies, and no PSD break is significantly detected,
suggesting an energy-dependent evolution not unlike that exhibited by
several Galactic black hole systems.
\end{abstract}



\section{Introduction}

In Seyfert Active Galactic Nuclei (AGNs), the observation of ubiquitous
rapid variability of the X-ray continuum emission
has established that these X-rays must originate very close to the central supermassive black hole,
in the central tens of $R_{\rm Sch}$ ($\equiv 2GM_{\rm BH}/c^2$), and
in the region of extreme gravity.  The X-rays are generally though to originate in some type of optically-thin,
hot ($10^{\sim9}$ K) corona; plausible geometrical configurations include, but are not limited to,
the hot, ionized outer layers of the optically-thick, geometrically-thin accretion disk which feeds the black hole
(Nayakshin et al.\ 2000),
part of a central, advection-dominated, radiatively-inefficient accretion flow (e.g., Narayan \& Yi 1994)
or the base of an outflowing jet (Markoff et al.\ 2005), 

A common tool to quantify the aperiodic variability as a function of temporal frequency
is the power spectral density (PSD) function. The {\it Rossi X-ray Timing Explorer} (\textit{RXTE}), launched in 1995,
has allowed multi-time scale monitoring campaigns to be executed, yielding 
high-quality, broadband X-ray PSDs probing X-ray variability on temporal frequencies from
$\sim 10^{-4}$ to $\sim 10^{-8}$ Hz (time scales from
hours to years). These PSDs commonly show "breaks" at temporal frequencies $f_{\rm b}$, 
with PSD power-law slopes breaking from $\sim$--2 to $\sim$--1 above and below $f_{\rm b}$, respectively.
Values of $f_{\rm b}$ are usually in the range $10^{-6}$ to $10^{-3}$ Hz, i.e.,
break time scales $T_{\rm b}$ ($\equiv 1/f_{\rm b}$) are usually in the range $\sim$0.01 to a few days
(e.g., Edelson \& Nandra 1999; Uttley, M$^{\rm c}$Hardy \& Papadakis 2002, hereafter U02;
Markowitz et al.\ 2003, hereafter M03; 
M$^{\rm c}$Hardy et al.\ 2004, 2005, 2007; 
Markowitz et al.\ 2007; Porquet et al.\ 2007; Chatterjee et al.\ 2009).   
These papers also demonstrated that break frequencies were consistent with
a correlation between $T_{\rm b}$ and black hole masses $\mbh$,
estimated by optical measurements such as reverberation mapping (e.g., Peterson et al.\ 2004) or via
host galaxy stellar velocity dispersion $\sigma_*$ (e.g., 
Merritt \& Ferrarese 2001, Tremaine et al.\ 2002). The correlation extrapolates down to 
stellar-mass, Galactic black hole X-ray binary (BH XRB) systems, and, combined with
the remarkable similarity in PSD shapes between Seyferts and BH XRBs, strongly
supports the notion of identical variability mechanisms operating
in both classes of objects.

However, as first suggested by M$^{\rm c}$Hardy et al.\ (2004), there is an additional dependence on the accretion
relative to Eddington $\dot{m} \equiv L_{\rm Bol}/L_{\rm Edd}$, such that for a given 
$\mbh$, $T_{\rm b}$ decreases as $\dot{m}$ increases.
This "fundamental plane" between $T_{\rm b}$, $\mbh$ and $\dot{m}$ 
signifies a sort of "unification" across supermassive and stellar-mass 
accreting black hole systems, and has been 
quantified in an empirically-derived relation by M$^{\rm c}$Hardy et al.\ (2006) from a sample of 
Seyferts and BH XRB X-ray PSDs.

In this paper, we present the broadband X-ray PSD of the Seyfert AGN
NGC 7469 and present evidence for a PSD break at time scale of $\lesssim$ 10 days.
As this source's black hole mass is well known both from reverberation mapping 
the optical broad line region (from an International AGN Watch monitoring program, e.g.,
Peterson et al.\ 2004) and from the $\mbh$--$\sigma_*$  relation, we can 
test if the break is consistent with the M$^{\rm c}$Hardy et al.\ (2006) empirical relation. 
Nandra \& Papadakis (2001; hereafter NP01) used an \xte\ intensive monitoring campaign
on NGC 7469 to measure preliminary high temporal frequency PSDs, 
over the temporal frequency range $10^{-6}$ to $10^{-3}$ Hz. They did not
find evidence for a break in the PSD over those temporal frequencies, 
but they claimed that the PSD slope flattened
as photon energy increased. The PSDs we present here
cover a wider temporal frequency range, and so we can revisit this claim. 
The rest of this paper is organized as follows: the light curve sampling and data reduction
are described in Section 2. 
The PSD measurement and fit results are presented in Section 3. The results are discussed in
Section 4, and a summary of our main conclusions is given in Section 5.
Appendix A contains the results of spectral fits to summed spectra derived from the
\textit{RXTE} campaigns.
Appendix B describes the ``surrogate'' Monte Carlo method of Press et al.\ (1992)
used to derive estimates of
PSD model parameter uncertainties.


\section{Observing Strategy, Observations, and Data Reduction}

Most of the high dynamic range Seyfert PSDs have been based on similar observing strategies:
regularly sampling the source flux over multiple time scales such that the resulting set of
light curves yields individual PSD segments covering complementary ranges in temporal frequency
(e.g., Edelson \& Nandra 1999)\footnote{Exceptions can occur for objects whose PSD break frequencies
lie at relatively high temporal frequencies, $10^{\sim -4}$ Hz;
for those objects, a PSD derived from an uninterrupted \textit{XMM-Newton} long-look of $\sim$ a hundred ks in duration
can reveal the break, e.g., Vaughan \& Fabian (2003) and Porquet et al.\ (2007).}. We follow the same strategy for NGC 7469.

We monitored the source with {\it RXTE} once every 4.27 days
(64 satellite orbits) for a duration of 6.3 years, from 2003 April 8 to 2009 July 15
(Modified Julian Day [MJD] 52737--55027;
observation identifiers 80152-05-*, 90154-02-*, 91138-02-*, 92108-04-*, 93144-06-*, and 94144-06-*).
This sampling, which probes variability on 
time scales from $\sim$ a week to a few years, is henceforth called
``long-term'' sampling. Each observation lasted approximately 1 ks.
There were six gaps due to satellite sun-angle viewing constraints
in mid-February to early April of each year; each gap was $\sim50$ days long.
For this light curve, the average deviation from ephemeris was 0.23 days (5$\%$).
More intensive monitoring with {\it RXTE} was done to probe
variability on time scales from a few hours to a month 
(``medium-term'' sampling). 
{\it RXTE} observed NGC 7469 from 1996 June 10 at 00:55 UT until 1996 July 12 at 00:21 UT
(MJD 50244.0--50276.0; observation identifiers 10315-01-*), obtaining 1-ks snapshots
every orbit (6 ks) for approximately the first half of the campaign, followed by
2-ks snapshots every other orbit (13 ks). 
Finally, we used the light curves from two continuous \textit{XMM-Newton} observations
to quantify variability on time scales from $\sim$ 1 hour to $\sim$ 1 day.

For the \textit{RXTE} sampling, 
we used data from the Proportional Counter Array (PCA; Swank 1998, Jahoda et al.\ 2006);
extraction of background-subtracted light curves followed standard extraction procedures
(we refer the reader to, e.g., M03 for additional details) and HEASOFT version 6.7 software.
PCA STANDARD-2 data were collected from proportional counter units (PCUs) 
0, 1, and 2 for the medium-term data (as PCUs 3 and 4 frequently exhibit breakdown during on-source time)
and PCU 2 only for the long-term data (as by 1998, PCU 1 also began to exhibit repeated breakdown during on-source time, and PCU 0 lost its propane veto layer in 2000 May). PCU 2 is thus the best-calibrated PCU.
We used standard screening criteria, including rejecting
data gathered within 20 minutes of satellite passage through the South Atlantic Anomaly (SAA).
The PCA background was estimated using the ``L7-240'' background models, appropriate for faint sources.

We extracted flux light curves;
spectral fitting for each observation was done using \textsc{xspec} version 12.5.1n,
assuming a Galactic column of $4.5 \times 10^{20}$ cm$^{-2}$ (Kalberla et al.\ 2005), 
the abundances of Wilms et al.\ (2000), and the cross sections of Verner et al.\ (1996).
Response files were generated for each separate observation using \textsc{pcarsp} version 7.10 to
account for the gradual hardening of the PCA response due to the gradual leak of xenon gas into the propane layer 
in each PCU. 
NGC 7469 contains a complex warm absorber system lying along the line of sight
to the continuum source, but these absorbers do not significantly impact the spectrum above 2--3 keV 
($\lesssim 1\%$ of continuum flux absorbed, e.g., Blustin et al.\ 2007) and are ignored in our modeling.
Errors on each flux point were derived from the standard error of 
16 s count rate light curve bins within each observation.
For the long-term light curve, the total number of data points after screening was
538, with 85 (15.8$\%$) missing due to, e.g., sun-angle constraints
or screening. The medium-term light curve, binned to a time scale of 6 ks, 
contained 479 pts, with 73 pts (15.2$\%$) missing due to screening or
time-critical observations of other sources; most of the missing points were in the second half of the
campaign as the sampling time was decreased (see NP01).
We extracted light curves in the 2--10, 2--5, 5--10, and 10--20 keV bands for both the
long- and medium-term time scales.

For the short-term sampling, we used public archive data 
from two continuous \textit{XMM-Newton} observations
(hereafter referred to as ``short1'' and ``short2'').
\textit{XMM-Newton} observed NGC 7469 starting at
2004 November 30 at 21:12 UT (revolution 912) for a duration of 85 ks,
and again starting at 2004 December 3 at 01:28 (revolution 913) for a duration of 79 ks.
We downloaded the pipeline processed data (version 6.6.0) from the HEASARC archive, and used
data from the European Photon Imaging Camera (EPIC) pn, which observed in Small Window mode
with the medium filter. No MOS data were used as the MOS cameras were in Timing Uncompressed mode;
further details of the observation can be found in Blustin et al.\ (2007).
Using \textsc{xselect} version 2.4a, we extracted 
source photons from a circular region of radius 40$\arcsec$ for each observation; backgrounds were
extracted from circular regions of identical size, centered $\sim$3$\arcmin$ away. 
We searched for background flares by inspecting the 10--13 keV pn light curves, finding none.
We extracted light curves in the 2--10, 2--5, and 5--10 keV bands
binned to 2000 s; variability at shorter time scales was 
dominated by Poisson noise (see below). 

Light curve sampling parameters, including 
mean net source and background count rates and average observed fluxes
for all light curves are listed in Table 1.
Also listed in Table 1 are fractional variability amplitudes $F_{\rm var}$ (see Vaughan et al.\ 2003 for
definition of $F_{\rm var}$ and its error) for each light curve. 

\textit{XMM-Newton} does not have coverage $>$ 10 keV, so we do not have
short-term sampling in the 10--20 keV range. We investigated if the longest uninterrupted 
10--20 keV PCA light curves during the longest duration observations from
the 1996 intensive monitoring could be used. However, we found the resulting 
PSDs to be dominated by Poisson noise at temporal frequencies above $\sim10^{-3.4}$ Hz, typically,
and not highly suitable for PSD analysis.
We also examined 20--40 keV light curves from the High Energy X-ray Timing Experiment (HEXTE) detectors
aboard \textit{RXTE} on all three time scales, but found the resulting PSDs to 
generally be dominated by Poisson noise over most temporal frequencies of interest,
so we do not investigate PSDs in that energy band.

2--10 keV light curves are displayed in Figure 1 for all time scales.


\section{PSD Measurement and Fit Results}

The PSD measurement procedure is summarized briefly here; the reader is referred to U02 or M03 for details.
Light curves were linearly interpolated across gaps.
Periodograms were constructed using a discrete Fourier transform (e.g., Oppenheim \& Shafer 1975)
and using the normalization of Miyamoto et al.\ (1991) and van der Klis (1997). 
Following Papadakis \& Lawrence (1993) and Vaughan (2005), 
the periodogram was logarithmically binned 
every factor of 1.4 in $f$ (0.15 in the logarithm)
to produce the observed PSD $P(f)$; the two lowest
temporal frequency bins were widened to accommodate three periodogram points, yielding 13, 12, 6, and 6 binned PSD
points for the long, medium, short1, and short2 light curves, respectively.
The observed PSDs are plotted in Figure 2 for the 2--10 keV band and in Figure 3 for the
sub-bands.

The $\sim$5 lowest temporal frequency
bins in each individual PSD typically contained less than 15 periodogram points, precluding us from
assigning normal errors.
To estimate proper errors on each binned PSD point, to account for the PSD measurement distortion effects of
red noise leak and aliasing, and to account for the effect of Poisson noise, we use the
Monte Carlo procedure outlined by U02 (based on
Done et al.\ 1992).  The vast majority of AGN X-ray PSD analyses conducted since 2002 
have used this procedure. For each PSD model shape tested, an average model $\overline{P_{{\rm sim}}(f)}$ is calculated
based on simulated PSDs; $\overline{P_{{\rm sim}}(f)}$ accounts for the distortion effects and has errors assigned based on the
rms spread of the individual simulated PSDs within each temporal frequency bin.
For each model, the value of the test statistic 
$\chi^2_\textrm{dist}$ between $P_{{\rm sim}}(f)$ and the observed PSD is  
compared to an empirical distribution of simulated $\chi^2_\textrm{dist}$ values.
The ``rejection probability'' $R$ is a goodness of fit measure defined as 
the percentile of simulated $\xsqdist$ values 
exceeded by the value of the observed $\xsqdist$;
below we list the likelihood of acceptance $L \equiv 1 - R$.

The constant level of power due to Poisson noise is not subtracted from the data,
but instead is added to each model discussed below. 
For the long- and medium-term PSDs, the power due to Poisson noise is estimated using $P_{\rm Psn} = 2(\mu+B)/\mu^2$,
where $\mu$ and $B$ are the total net and background count rates,
respectively; as these light curves are non-continuous light curves we multiply our estimate of 
$P_{\rm Psn}$ by the average value of the ratio $\Delta$$T_{\rm samp}$, the average sampling time,
to the average exposure time per snapshot. 
For each short-term PSD, we used light curves binned to 300 s to 
measure the PSD out to $10^{-2.8}$ Hz, well into the temporal frequency range
dominated by Poisson noise (above 10$^{-3.6}$ Hz, typically).
A best fit to all binned PSD points in the range 10$^{-3.6}$ to $10^{-2.8}$ Hz yielded
the $P_{\rm Psn}$ values listed in Table 1.

Given that there is no overlap in time between the long-, medium-, and short-term 
light curves, we implicitly assume that the intrinsic variability process
is only weakly non-stationary
over 13 years. That is, we assume that the intrinsic PSD has remained constant in both shape and normalization for a given energy band,
and the expectation value of $F_\textrm{var}$ for a given energy band also remains constant.
This is a reasonable assumption: 
PSDs in BH XRBs tend to display significant changes in the shape or normalization of the
components comprising the PSD (e.g., Lorentzians and/or power-laws) on time scales of $\sim$ a day or longer (e.g., Pottschmidt et al.\ 2003;  Remillard \& McClintock 2006);
scaling with black hole mass and luminosity, AGN light curves may be expected to display strong
non-stationarity, with significant changes in the observed 
PSD, on time scales of at least centuries to millennia.

As the long-term monitoring spans a six-year duration, we can test the assumption of
weak non-stationarity over this duration by splitting the long-term light curve
in half (before and after MJD 53882), calculating the periodogram for each half, and 
using the procedure outlined by Papadakis \& Lawrence (1995)
to determine if the periodograms are consistent.
In this method, a statistic $S$ is calculated 
based on the sum of the differences in power at each temporal frequency
(see Appendix A of Papadakis \& Lawrence 1995 for the definition of $S$); for
consistent periodograms, $S$ has zero mean and a variance of 1.
For the periodograms derived from each half of the 2--10, 2--5, 5--10, and 10--20 keV light curves, we
calculate $S$ values of +0.08, --1.08, +0.20, and --1.10, respectively,
consistent with the notion that at each energy range studied, the PSD
has remained constant in shape and normalization between 2003 and 2009. 

The procedure of Papadakis \& Lawrence (1995) cannot be used to directly compare
the long- and medium-term PSDs. However, in BH XRBs, a change in the PSD is commonly 
accompanied by a significant change in the energy spectrum.
We thus performed fits to summed energy spectra
derived from the long- and medium-term \textit{RXTE} monitoring.
These results are presented in Appendix A and
demonstrate that the energy spectra are highly similar in form on all time scales
and are thus consistent with the notion of only weak non-stationarity
between 1996 and 2009.

\subsection{PSD Model Fits}

The PSDs of BH XRBs are usually of sufficient quality to model multiple components, including
Lorentzians and quasi-periodic oscillations (e.g., Remillard \& McClintock 2006).
However, the temporal frequency resolution of AGN PSDs usually means that simple
unbroken or singly-broken PSD model shapes provide an adequate fit (the double-peaked
profile of Ark 564 measured by M$^{\rm c}$Hardy et al.\ 2007 is a notable exception).
For NGC 7469, we tested three PSD model shapes: an unbroken power-law,
a broken power-law consisting of a sharp break, and a
broken power-law with a gradual bend connecting the high- and low-frequency portions.

The unbroken power-law model was of the form $P(f) = A_0 (f/f_0)^{-\alpha}$, 
where $\alpha$ is the power-law slope
and the normalization $A_0$ is the PSD amplitude at $f_0$, arbitrary chosen to be $10^{-6}$ Hz.
We stepped through $\alpha$ in increments of 0.02, testing the
range of slopes 0.5 to 2.5, and using $N_{\rm trial}$ = 200 simulations each time 
to calculate $\overline{P_{{\rm sim}}(f)}$.
Best-fit model parameters, likelihoods of acceptance $L_{\rm unbr}$, and values of observed
$\chi^2_{\rm dist}$/dof are listed in Table 2. 
The data--model residuals are plotted in Figure 2(c) for the 2--10 keV band and Figure 3(c) for the
sub-bands.

In Appendix B, we discuss various methods for estimating the (one-dimensional)
confidence ranges for each fitted model parameter for this and all subsequent PSD models. 
There, we discuss implementation of
the ``surrogate'' Monte Carlo method of Press et al.\ (1992);
the resulting confidence ranges are reported in parentheses in Table 2 for the unbroken
power-law model and in Table 3 for the broken-power models discussed below.

The values of $\chi^2_{\rm dist}$/dof are generally poor for the 2--10, 2--5, and 5--10 keV PSDs,
and the likelihoods of acceptance are low ($<$0.3$\%$).
The 10--20 keV PSD, lacking short-term sampling, covers less dynamic range 
and the likelihood of acceptance is much higher.
For the lower-energy bands, the residuals plotted in Figures 2(c) and 3(c) suggest that a more
complex PSD model shape is appropriate.

To test for the presence of a PSD break, we employed a power-law model with a sharp break
of the form
\[P(f)= \left\{ \begin{array}{ll}
                              A_1(f/f_{\rm b})^{- \alpha_{\rm lo}},  & f \le f_{\rm b} \\
                              A_1(f/f_{\rm b})^{- \alpha_{\rm hi}},   & f > f_{\rm b} \end{array}
\right. \]
where the normalization $A_1$ is the PSD 
amplitude at the break frequency $f_{\rm b}$, and
--$\alpha_{\rm lo}$ and --$\alpha_{\rm hi}$ are the low- and high-frequency power law slopes, respectively,
with the constraint $\alpha_{\rm lo} < \alpha_{\rm hi}$.

Break frequencies were tested in the log from --7.4 to --4.9
in increments of 0.1.  $\alpha_{\rm hi}$ and $\alpha_{\rm lo}$ 
were both tested in increments of 0.1, over the ranges 1.0--3.2 and 0.0--2.0, respectively.
One hundred simulated PSDs were used to determine $\overline{P_{{\rm sim}}(f)}$.
The best-fit model parameters, along with likelihoods of acceptance $L_{\rm brkn}$, are listed in Table 3. 
Errors listed are for one interesting parameter and
were determined assuming that other parameters (except for $A_1$) were fixed.
Data--model residuals are plotted in Figure 2(d) for the 2--10 keV PSD and
in Figure 3(d) for the sub-band PSDs.
Figure 4 shows contour plots of $\alpha_{\rm hi}$ versus $f_{\rm b}$ for these three PSDs
at the respective best-fit values of $\alpha_{\rm lo}$.
Best-fit values of $f_{\rm b}$ lie near $2 \times 10^{-6}$ Hz for all PSDs, with best-fit values of 
$\alpha_{\rm lo}$ close to 1.0. Excluding the 10--20 keV PSD, best-fit values of
$\alpha_{\rm hi}$ are found to be 1.8--1.9.


We can use the ratio of the likelihoods of acceptance $L_{\rm brkn}/L_{\rm unbr}$ between the broken
and unbroken power law model fits to establish that incorporating a break into the model 
yields a significant improvement. The likelihood ratio test 
$D \equiv -2$ln($L_{\rm unbr}/L_{\rm brkn}$)
is distributed as a $\chi^2$ distribution with $n$ degrees of freedom, where $n$ is the difference in degrees of freedom between
the two models, here equal to 2.
For the 2--10, 2--5, and 5--10 keV PSDs, 
values of $L_{\rm brkn}/L_{\rm unbr}$ are in the range 17--67 and values of $D$ span 5.7--8.4,
indicating an improvement in fit with respect to
the null hypothesis model (no break required) at confidence levels spanning 94.0--98.5$\%$.
However, for the 10--20 keV PSD, the improvement in fit
when adding a break to the model is not significant, with $L_{\rm brkn}/L_{\rm unbr} = 2.4$ and $D=1.72$,
an improvement in fit over the null hypothesis model at only 58$\%$ confidence.
Combined with the fact that the best-fit values of 
$\alpha_{\rm lo}$ and $\alpha_{\rm hi}$ are consistent with other, this signifies that a break
has not been robustly detected in the 10--20 keV PSD.




We also tested a more slowly-bending PSD 
model of the form $P(f) = (A_1 f^{-\alpha_{\rm lo}})/( ( 1 + f/f_{\rm b})^{(\alpha_{\rm hi} - \alpha_{\rm lo})}$,
testing the same range and increments of $f_{\rm b}$, $\alpha_{\rm hi}$ and $\alpha_{\rm lo}$ 
as for the sharply-broken power-law model.
The best-fit model parameters and likelihoods of acceptance $L_{\rm slow}$ are listed in Table 3,
with data--model residuals plotted in Figure 2(e) for the 2--10 keV PSD and Figure 3(e) for the sub-band PSDs.
Contour plots of $\alpha_{\rm hi}$ versus $f_{\rm b}$ at the respective best-fit values of $\alpha_{\rm lo}$
are shown in Figure 4.
As with the sharply-broken PSD model fits, the improvement in fit when adding a break
is significant: for the 2--10, 2--5 and 5--10 keV PSDs, 
$L_{\rm brkn}/L_{\rm unbr}$ spans 29--75, $D$ spans 6.8--8.6, and the improvement with
respect to the unbroken power-law is at confidence levels spanning 96.5--98.6$\%$. The evidence for a break in the 
10--20 keV PSD is again unconvincing, with consistent best-fit values of 
$\alpha_{\rm lo}$ and $\alpha_{\rm hi}$ and with $L_{\rm brkn}/L_{\rm unbr} = 2.4$  
and $D=1.77$, an improvement over the null hypothesis model at only 59$\%$ confidence.
As both classes of broken power-law models fit roughly similarly and yield significant evidence
for a PSD break in the 2--10, 2--5 and 5--10 keV bands, we henceforth treat both models equally in the paper.

\subsection{The PSD as a function of photon energy}

We now examine the behavior in PSD parameters with photon energy; specifically,
we focus on the apparent lack of a significant detection of a break in the 10--20 keV PSD.
To quantify differences between the observed
behavior of the 10--20 keV PSD and that of the PSDs at lower energies,
we define $\Delta\alpha$ as the difference between the power-law slopes below and above the best-fit PSD break,
using the parameter errors as listed in Table 3. 
For the singly-broken power-law model,
the average of $\Delta\alpha$ for the 2--5 and 5--10 keV PSDs is $1.0 \pm 0.3$,
while for the 10--20 keV PSD, $\Delta\alpha$ is $0.1 \pm 0.2$.
The difference in $\Delta\alpha$ from below 10 keV to above 10 keV 
suggests that the absence of a break in the 10--20 keV PSD
and the appearance of a break in the $<$ 10 keV PSDs
are significant at approximately the  4$\sigma$ confidence level.
Similarly, for the slowly-bending model,
the average of $\Delta\alpha$ for the 2--5 and 5--10 keV PSDs is $ 1.4 \pm 0.3$,
while for the 10--20 keV PSD, $\Delta\alpha$ is  $0.2\pm0.3$;
the difference between these $\Delta\alpha$ values also suggests
that the observed difference in PSD behavior is significant at the $\sim$4$\sigma$ confidence level.

Figure 5 shows an overplot of the observed PSDs in each band in data-space
and in ``model-space'', along with the best-fitting sharply-broken and slowly-bending power-law models.  
Above $\sim10^{-6}$ Hz, the 10--20 keV PSD is much flatter than the 2--10, 2--5, and 5--10 keV PSDs.
Similar behavior at high temporal frequencies was reported by NP01.
They constructed PSDs from the medium-term sampling, covering temporal
frequencies from $10^{-5.8}$ to $10^{-4.1}$ Hz 
and used uninterrupted PCA light curves with a time resolution of 16 s
to probe $10^{-3.5}$ Hz to $10^{-1.5}$ Hz in the 2--4, 4--10, and 10--15 keV PSDs.
NP01 fixed $P_{\rm Psn}$ to the expected noise power level instead of leaving it as a 
free parameter, but this yielded a good fit at high temporal frequencies
(as stated earlier, we do not use PCA to measure the 10--20 keV PSD because $P_{\rm Psn}$ 
dominates). However, NP01 did not take into account PSD distortion measurement effects
(the effects of which will have an energy dependence if the
PSD shape itself is energy-dependent). Nonetheless, the current work and NP01 both
find a similar result: the 10--20 keV PSD is much flatter at higher temporal frequencies.
While the 10--20 keV band lacks short-term sampling,
this behavior cannot be due to an effect of sampling on the medium or long time scales, as \textit{RXTE}
sampled all four bands equally.



We also compared the 2--5 and 5--10 keV bands to search for any PSD evolution in
energy between those two bands. We first tested for any dependence of $f_{\rm b}$ with energy
assuming a universal PSD shape whose power-law slopes
are independent of energy.
Assuming the best-fit sharply-broken model to the 2--10 keV PSD,
we find consistent values for $f_{\rm b}$ for the 2--5 and 5--10 keV bands, with log($f_{\rm b}$,Hz) = $-5.6 \pm 0.3$
for each band. We find log($f_{\rm b}$,Hz) = --($6.0 \pm 0.4$) and --($6.0\pm0.3$) for the 2--5 and 5--10 keV bands, respectively, 
using the slowly-bending model.
We then tested for an energy dependence of $\alpha_{\rm hi}$ assuming that $f_{\rm b}$ is energy-independent. 
For a sharply-broken model with log($f_{\rm b}$,Hz) fixed at --5.7, the values of $\alpha_{\rm hi}$ are consistent for
the 2--5 and 5--10 keV bands: $1.8 \pm 0.2$ and $1.8^{+0.1}_{-0.2}$, respectively
($1.8 \pm 0.4$ for each band for the slowly-bending model assuming log($f_{\rm b}$,Hz) is fixed at --6.0).
In conclusion, there is no significant evidence for PSD evolution in energy between the 2--5 and 5--10 keV bands.



\section{Discussion}


\subsection{NGC 7469's Place in the $\Tb$--$\mbh$--$\dot{m}$ Plane}

The primary result of this paper is that we detect a turnover in the 2--10 keV PSD of NGC 7469 at a
temporal frequency of $f_{\rm b}= 2.0^{+3.0}_{-0.8} \times 10^{-6}$ Hz or $1.0^{+3.0}_{-0.6} \times 10^{-6}$ Hz  
for the sharply- or slowly-bending power-law model, respectively (using the 68$\%$ confidence limits from the P92 method);
these frequencies correspond to turnover time scales $5.8 \pm 3.5$ days or $11.6^{+17.5}_{-8.7}$ days, respectively.

We first determine if the measured PSD turnovers are consistent with the empirical relation  
between $\Tb$, $\mbh$, and $\dot{m}$ quantified by M$^{\rm c}$Hardy et al.\ (2006).
As noted above, NGC 7469's black hole mass is well studied.
We use the reverberation-mapped mass from Vestergaard \& Peterson (2006), who provide
a "calibrated" mass estimate based on H$\beta$ width and optical luminosity measurements of
$\mbh = 3.34^{+0.69}_{-0.68} \times 10^7 \Msun$.
Our conclusions below do not change if we use the mass estimate from
the $\mbh$--$\sigma_*$ relation: $\sigma_* = 131 \pm 5$ km s$^{-1}$ (Nelson et al.\ 2004), and
using the $\mbh$--$\sigma_*$ relation of Tremaine et al.\ (2002)
yields $\mbh = 2.46^{+0.44}_{-0.34} \times 10^7 \Msun$. 

For an estimate of the bolometric luminosity, we use the value from
Vasudevan \& Fabian (2009) based on \textit{XMM-Newton} EPIC/Optical Monitor spectral energy distribution 
fitting, $L_{\rm bol} = 6 \times 10^{44}$ erg s$^{-1}$.
An independent estimate of $L_{\rm Bol}$ can be based on the 
\textit{RXTE} monitoring data: a model fit jointly to the summed spectrum 
\textit{RXTE} PCA and HEXTE spectra constructed from all available \textit{RXTE} observations of NGC 7469
(Rivers et al., ApJS, submitted) indicates that 
the modeled average unabsorbed 2--10 keV flux is $3.1 \times 10^{-11}$ erg cm$^{-2}$ s$^{-1}$.
Using a luminosity distance of 62.7 Mpc (from the NED database, using the reference frame defined by the 3K cosmic microwave
background radiation), 
the 2--10 keV luminosity $L_{2-10}$ is $1.7 \times 10^{43}$ erg s$^{-1}$. 
Using Marconi et al.\ (2004), $L_{\rm bol} \sim 22 L_{2-10} = 3.7 \times 10^{44}$ erg s$^{-1}$,
very close to the value from Vasudevan \& Fabian (2009), which we use below.
  
Using the best-fit values of the coefficients in the
M$^{\rm c}$Hardy et al.\ (2006) relation, 
log($T_{\rm b,pred}$) = 2.10log($\mbh$/($10^6 \Msun$)) -- 0.98log($L_{\rm bol}/(10^{44} {\rm erg~s^{-1}}$) -- 2.32,
with $T_{\rm b,pred}$ in units of days, and taking into the account the uncertainty in $\mbh$ listed above,
we obtain $T_{\rm b,pred} = 1.25^{+0.60}_{-0.48}$ days. 
Taking into account the uncertainties on
the coefficients in the M$^{\rm c}$Hardy et al.\ (2006) relation, 
$2.10\pm0.15$, $0.98\pm0.15$, and $2.32\pm0.2$, respectively, we can obtain
$T_{\rm b,pred}$ values from 0.2 to 9.3 days, meaning the predicted
value is consistent with both the observed sharply-broken and slowly-bending PSD break time scales.

\subsection{Corresponding Physical Time Scales}


Detailed discussions on the likely physical mechanisms responsible for the
PSD turnovers have appeared in many of the PSD papers cited above; here,
we provide only a brief review.

We assume that the bulk of the X-ray emission originates 10 $R_{\rm Sch}$ from the black hole.
The Keplerian orbital time scale $t_{\rm orb}$ at this radius for a $3 \times 10^7 \Msun$ black hole is 
1 day.  Following, e.g., Treves et al.\ (1988), the thermal time scale $t_{\rm th}$ is roughly $t_{\rm orb} / \alpha$, where
$\alpha$ is the accretion disk viscosity parameter. For values of $\alpha$ of $\sim 0.1$--$0.2$,
$t_{\rm th}$ will be $\sim 5-10$ days, consistent with the observed PSD break time scales in NGC 7469. 

One variability model that has had a measure of success in explaining 
the observed variability properties of Seyferts and BH XRBs,
including PSD shapes, temporal frequency-dependent lags, and the rms--flux relation, involves
inwardly-propagating fluctuations in the local mass accretion rate (Lyubarskii 1997;
Kotov, Churazov \& Gilfanov 2001; Ar\'{e}valo \& Uttley 2006).
In this model, the fluctuations originate across a range of annuli in the disk and travel
inward toward the central X-ray source, eventually modifying the X-ray continuum emission.
The PSD breaks could correspond to the local viscosity time scale $t_{\rm visc}$ at the outer radius of X-ray emission.
$t_{\rm visc} = t_{\rm th} / (H/R)^2$, where $H/R$ is the ratio of the disk scale height
to the radius (e.g, Treves et al.\ 1988). 
For a geometrically thin disk surrounding 
a $3 \times 10^7 \Msun$ black hole, with $H/R = 1/100$, and 
with $\alpha \sim 0.1-0.2$, $t_{\rm visc} \sim  5-10 \times 10^4$ days,
far too long to be associated with the observed PSD breaks. However, for a geometrically thick disk where
$H/R$ approaches 1, $t_{\rm visc}$ approaches the thermal time scale and can match the
observed PSD time scale for NGC 7469.
However, Ar\'{e}valo \& Uttley (2006) caution that if the X-ray emission region is radially extended, 
one can still get a bend in the PSD due to the radial distribution in variability fluctuations,
and the PSD bend does not have to correspond to a singular characteristic time scale.





\subsection{Energy Dependence of the PSD}

Rapid variability in the Compton hump is not a likely cause for the observed energy-dependence of the PSD.
The Compton hump contributes only $\sim$20--30$\%$ of the total 10--20 keV emission (see Table 4 in Appendix A), and so
such extreme variability in the total 10--20 keV emission
would require the bulk of the Compton-thick reflecting material to originate within 
light-hours of the X-ray continuum source and to be responding to continuum variations much larger than those we observe.
Nandra et al.\ (2000) explored variability in the absolute normalization $A_{\rm ref}$ of the Compton hump down to time scales of 1 day,
but found the observed variability in $A_{\rm ref}$ consistent with being due to model degeneracy between $\Gamma$ and $A_{\rm ref}$.
Finally, Papadakis, Nandra \& Kazanas (2001) demonstrated that the emission in 2--10 and 10--15 keV bands show high coherence over $10^{-5.5}$ to $10^{-4}$ Hz, suggesting a common variability mechanism for both bands.

Flattening of the power-law slope at temporal frequencies above the break
with increasing energy has been claimed for some PSDs
published previously, such as Mkn 766 (Vaughan \& Fabian 2003, Markowitz et al.\ 2007)
and MCG--6-30-15 (Vaughan, Fabian \& Nandra 2003); $\alpha_{\rm hi}$ was
typically 2.5 for bandpasses centered near 0.5 keV and 2.1 near 5 keV.
Moreover, modeling the broadband PSD of Ark 564 using a double-Lorentzian profile,
M$^{\rm c}$Hardy et al.\ (2007) found the normalization of the 
higher-temporal frequency Lorentzian to increase by a factor of 1.5
from the 0.6--2.0 to the 2--10 keV bands.

Interestingly, evolution in PSD shape as a function of photon energy is not 
uncommonly observed in BH XRBs, and is
commonly attributed to changes in the normalizations and/or peak temporal frequencies
of the Lorentzians components used to fit the PSD. Energy-dependent
PSD changes do not tend to be confined to any particular energy spectral state
(Done \& Gierli\'{n}ski 2005). For example, Kalemci et al.\ (2003, their Figure 5)  
studied the PSD of XTE J1650--500 during an outburst decay, modeling the PSD
using a single Lorentzian profile. They noted that the 
6--15 keV Lorentzian peaks at a temporal frequency
a factor of 3 higher than that in the 2--6 keV band,
and the normalization also increased by a factor of 3, yielding a much flatter
PSD above a temporal frequency of $\sim$3 Hz.
B\"{o}ck et al.\ (2009), modeling the 4.5--5.8 keV and 9.5--15 keV PSDs 
of an intermediate state of Cyg X-1 using a double-Lorentzian profile,
noted that the lower-temporal frequency Lorentzian decreased in normalization
toward the higher-energy band, while the normalization of the other Lorentzian remained
the same, yielding a flatter overall PSD above $\sim$3 Hz in the higher-energy band.
Similar changes between PSDs measured over the 2--4 and 15--71 keV bands were seen 
across several spectral states of Cyg X-1 by Pottschmidt et al.\ (2006).
In addition, many quasi-periodic oscillations are 
detected only above a threshold photon energy (e.g., 
Strohmayer 2001a, 2001b; Montanari et al.\ 2009). 



Though we have modeled the PSD of NGC 7469 using a simple singly-broken power-law,
we cannot rule out the possibility that multiple broadband components may exist in this PSD, and, speculatively, 
the observed PSD flattening with energy in NGC 7469 and other Seyferts 
could be due to changes in the relative normalizations of these components, as in Ark 564 (M$^{\rm c}$Hardy et al.\ 2007).
Another possibility is that the break time scale could increase dramatically from 
approximately $10^{-6}$ Hz in the 2--10 keV band to $10^{-4}$ Hz or greater in the 10--20 keV band. 
In the context of a model incorporating inwardly-propagating disk fluctuations,
if the bulk of the 10--20 keV emission originates at a smaller radius compared to the 2--10 keV emission 
(e.g., Kotov, Churazov \& Gilfanov 2001), then
a jump in the physical parameters of the disk with radius (e.g., if $H/R$ increases by 10) 
could cause vastly different values for $t_{\rm visc}$ at different radii of the disk.
A final possibility is that the break frequency in the PSD of NGC 7469 occurs at the same temporal frequency
in all bands, with the power-law slope flattening with photon energy.
This behavior is (at least qualitatively) consistent with 
the jet model of Giannios et al.\ (2004; see also Kylafis et al.\ 2008), wherein
the Comptonizing corona is identified with the base of a jet, 
and a PSD break corresponds to the Keplerian frequency at the radius of the jet base. 
PSD flattening above the break is attributed to having the temperature of jet decrease with increasing radius
and having the variability of the soft input photons governed by the radius at which they are emitted from the disk.


\section{Conclusions}    

We have presented the broadband X-ray PSD of the X-ray-typical Seyfert 1.2 NGC 7469.
Preliminary PSDs for this object were published by NP01 and based on
a 1996 month-long \textit{RXTE} campaign which quantified variability over temporal frequencies greater
than approximately 10$^{-6}$ Hz. We combined these data with 
sampling obtained from long-term monitoring with \textit{RXTE} spanning 2003--2009 
and two \textit{XMM-Newton} long-looks obtained in 2004.
The resulting high dynamic range of the PSD, $9 \times 10^{-9}$ to $2 \times 10^{-4}$ Hz,
allowed us to test simple unbroken and singly-broken power-law model shapes.
In the 2--10 keV PSD, we find significant evidence for a break at a best-fit temporal frequency
$ f_{\rm b} = 1-2 \times 10^{-6}$ Hz, depending on the form of the power-law break.
This corresponds to a time scale of $T_{\rm b} = 6-12$ days.
Given NGC 7469's well-constrained black hole mass of $3 \times 10^{7} \Msun$, the PSD break is consistent with the
empirical relation between $\mbh$, $L_{\rm bol}$ and $T_{\rm b}$ of M$^{\rm c}$Hardy et al.\ (2006) for Seyfert PSDs.

Our results include reliably derived confidence regions on the best-fit parameters
on models fit to the 2--10 keV PSD. We applied the ``surrogate'' method of Press et al.\ (1992), wherein 
one uses Monte Carlo simulations to create a large number of synthetic data sets based on the best-fit model,
and fits each data set to map out the probability distributions of model parameters.
The parameter confidence ranges listed in Tables 2 and 3 thus lack the
ambiguity inherent in parameter confidence regions derived from previously derived methods 
(Markowitz et al.\ 2003; Uttley et al.\ 2002),
though we emphasize that work is still ongoing in this area (e.g., Mueller \& Madejski 2009).

A break is not confirmed in the 10--20 keV band PSD, which is significantly flatter than the
2--10 keV PSD (and the 2--5 and 5--10 keV sub-bands), consistent with claims by NP01.
Including the current result for NGC 7469, evidence for energy-dependent PSD evolution in Seyferts has been accumulating,
and includes an energy-dependent change in the normalization of a Lorentzian component 
(as in Ark 564; M$^{\rm c}$Hardy et al.\ 2007), 
and detections of evolution in power-law slope at temporal frequencies above the break (e.g., Mkn 766, Vaughan \& Fabian 2003, Markowitz et al.\ 2007;
MCG--6-30-15, Vaughan, Fabian \& Nandra 2003).
The accumulated evidence suggests that energy-dependent evolution may not be uncommon in Seyfert PSDs and
may be similar to energy-dependent changes observed in the (higher-quality) PSDs of several BH XRBs, thereby
corroborating other timing-based observational links between Seyferts and BH XRBs 
(broadband PSD shapes and the scaling of PSD break frequencies with black hole mass and accretion rate
relative to Eddington; coherence; phase lags)
which support the notion of similar variability mechanisms at work
in both classes of objects.
For further progress, the community needs a large sample of AGN PSDs observed with the same quality as XRB PSDs in order to perform 
the same level of detailed model fitting of Lorentzian components, power-law components, etc.,
and determine if and how those components vary in energy.
The required PSDs would need adequately high temporal frequency resolution covering both a broad temporal frequency range
and a broad photon energy range, including obtaining low Poisson noise measurements above at least 10 keV
and covering temporal frequencies up to at least $\sim$10$^{-3}$ Hz.



\acknowledgements 
A.M.\ thanks the {\it RXTE} Science Operations staff, particularly the {\it RXTE}
schedulers for ensuring that the long-term monitoring observations were scheduled
so evenly all these years, thereby ensuring straightforward PSD measurement.
A.M.\ also thanks Martin Mueller, Katja Pottschmidt, and Moritz B\"{o}ck for useful discussions
which helped guide the manuscript. This work has made use of HEASARC online services, supported by
NASA/GSFC, and the NASA/IPAC Extragalactic Database,
operated by JPL/California Institute of Technology under
contract with NASA.

\appendix
\section{A. The X-ray Energy Spectrum of NGC 7469 from 1996 to 2009}

In this Appendix, we compare energy spectra
of NGC 7469 obtained during the long-, medium-, and short-term observations
to check for any evidence of significant spectral changes from 1996 to 2009 which might be associated with
strong non-stationarity in the underlying variability process.
As mentioned previously, in BH XRBs, significant changes in the shapes and normalizations of
the broadband PSD and its components are commonly accompanied by significant
changes in the energy spectrum.  For example, 
the ``low/hard'' spectral state commonly seen at relatively low
values of $\dot{m}$ (typically 0.01--0.04 for most BH XRBs; 
Maccarone 2003) includes a flat energy spectrum with power-law continuum
photon index $\Gamma$ $\sim$ 1.6--1.8, 
while in the thermally dominated ``high/soft'' state,
a strong soft thermal component with typical $k_{\rm B}T \sim 1$ keV emerges, 
the power-law continuum is steeper ($\Gamma \gtrsim 2$), 
and there is relatively weaker 2--10 keV variability (e.g., Remillard \& McClintock 2006).

As pointed out by, e.g., Done \& Gierli\'{n}ski (2005), 
directly analogizing Seyferts with BH XRB energy states
by comparing PSD and energy spectral properties 
is not straightforward; complications include 
the effects of having vastly different disk temperatures in the two classes of objects
and the fact on time scales less than years--decades, 
most Seyferts are persistent while many BH XRBs are transients.
However, we can examine the energy spectra of NGC 7469 during the 1996 and
2003--2009 \textit{RXTE} campaigns to check for any evidence of significant spectral changes
between the two non-overlapping campaigns.

We summed all PCA and HEXTE spectra from individual observations.
Extraction of spectra and PCA responses
followed standard procedures, and for details,
we refer the reader to Rivers, Markowitz \& Rothschild (ApJS, submitted), who 
present model fits to the 3--100 keV time-averaged 
spectrum, jointly fitting the total sum of all PCA and HEXTE data
over 3--30 and 15--100 keV, respectively. 
They found that a good fit was obtained
by using a model incorporating a power-law
continuum (with no evidence for a high-energy cutoff), 
an Fe K$\alpha$ emission line modeled with a Gaussian component,
and a Compton reflection hump modeled with a \textsc{pexrav} component
in \textsc{XSPEC} (Magdziarz \& Zdziarski 1995; inclination angle assumed to be 30$\degr$),
all modified by absorption by the Galactic column. 
0.5$\%$ systematics were added to the PCA spectrum.

We applied this model to the summed spectra over 1996 and 2003--2009, also jointly fitting the PCA
and HEXTE data over 3--30 and 15--100 keV, respectively. Good exposure times are listed in Table 4.
We allowed a constant factor for HEXTE/PCA cross-normalization;
best-fit values were near 0.8. Background level corrections were applied 
using \textsc{recorn} to adjust for imperfect predictions of background (PCA) and deadtime (PCA and HEXTE).
The best-fit value for the correction to PCA background in each spectral fit was
--0.9$\%$, and values for the corrections to HEXTE backgrounds were in the range --0.10$\%$ to +0.08$\%$.


We found good fits to both spectra using this model. Best-fit parameters are
listed in Table 4; errors correspond to $\Delta\chi^2 = 2.71$.  
The spectral data and residuals to the best-fit models are plotted in Figures 6 and 7
for the 1996 and 2003--2009 spectra, respectively.
Our best fit to the 1996 spectrum is consistent with 
that presented by Nandra et al.\ (2000). 
Similarly, there is rough agreement with the model fit to the 3--10 keV spectrum of the
\textit{XMM-Newton} spectra obtained in 2004, which Blustin et al.\ (2007) found to be
dominated by a power-law with $\Gamma = 1.81 \pm 0.01$. The lower value of
$\Gamma$ compared to the \textit{RXTE} fits is at least in part due to the fact that
Blustin et al.\ did not model a Compton reflection hump; omitting that component
from our fits causes the best-fit value of $\Gamma$ to fall to 1.80 
(1996) or 1.84 (2003--2009), in better agreement with Blustin et al.\ (2007).

Given the good agreement in spectral fit parameters, as well as in flux and
luminosity (implying that the average $\dot{m}$ did not change significantly between the
campaigns), we can conclude that there is no evidence for a 
spectral state change in NGC 7469 between 1996 and 2009, and that our
assumption of weak non-stationarity is valid.


\section{B. Notes on Estimation of Model Parameter Uncertainties}

In this Appendix, we briefly discuss methods for estimating the (one-dimensional)
confidence ranges for each fitted model parameter.

U02 defined a 90$\%$ confidence range for one interesting parameter
using the region of parameter space for which the fit rejection probability remained below 90$\%$; 
however, the sizes of the resulting confidence ranges depended critically on how close the 
rejection probability of a given best-fit model was to this value. For 
example, if a model had a rejection probability of say 88$\%$, the 90$\%$ confidence 
ranges were very small (see Figure 1 of Mueller \& Madejski 2009). 
In our Tables 2 and 3, the error on each fitted model parameter was derived
using a method introduced by M03. These errors 
correspond to values 1$\sigma$ above the rejection probability $R_{\rm unbr}$ for the best-fit
value on a Gaussian probability distribution. For 
example, if the best-fit model had $R_{\rm unbr}$ = 95.45$\%$
(2.0$\sigma$ on a Gaussian probability distribution), errors correspond to $R_{\rm unbr}$ = 99.73$\%$
(3.0$\sigma$). This estimate of parameter error, also used by e.g., 
Markowitz (2009), was used to alleviate the dependence of the size of the error range
on the value of the rejection probability of the best-fit model.
As noted by Mueller \& Madejski (2009), however, 
the likelihood of acceptance (rejection probability, equivalently) 
is a goodness of fit measure by its definition,
analogous to a $p$-value for $\chi^2$. For both the U02 and M03
methods, it is therefore not immediately clear to which confidence level the resulting derived 
error ranges really correspond. 


Following Press et al.\ (1992, hereafter P92), we thus undertook additional Monte Carlo simulations to 
map out the probability 
distributions of each of the fitted model 
parameters for a given best-fit model, assuming that the model
is indeed the best description of the data.
As described by P92, in this ``surrogate'' method, 
the best-fit model parameters to the observed PSD (e.g., listed in Tables 2 and 3), hereafter denoted $a_{\rm 0}(i)$
for a set of $i$ parameters ($\alpha$ for the unbroken power-law; $\alpha_{\rm lo}$, $\alpha_{\rm hi}$, and $f_{\rm b}$ for the broken power-law models)  
are used as ``surrogates'' for the true, underlying parameters $a_{\rm true}(i)$, and are used to generate $j$ 
sets of simulated data, each with best-fit parameters $a_{\rm sim}(i)$. 
The distributions of $a_{\rm sim}(i)$
map out the true parameter probability distributions, provided that certain conditions (discussed below) hold.  

We applied this method to the best-fit unbroken, sharply-broken and slowly-bending model fits 
to the observed 2--10 keV PSD. 
For each of these three models, we 
created $j=100$ simulated data sets of long, medium, short1 and short2 
light curves assuming the best-fit PSD model values $a_{\rm 0}(i)$ (listed in Tables 2 or 3),
and found the best-fit parameter values for each by repeating the Monte Carlo fit procedure.
For each data set, we stepped through the same ranges of parameters as explored above,
using the same increments, 
and selecting different random number seeds for each trial. 
We thus constructed distributions of 
the 100 best-fit values of $\alpha$ for the unbroken power-law model
and $\alpha_{\rm lo}$, $\alpha_{\rm hi}$ and $f_{\rm b}$ for the broken power-law models.
We used the San Diego Supercomputer Center's Triton Compute Cluster
to carry out this computationally intensive process, which 
required a total of 9100 processor hours to fully test all three model shapes.
The distributions $a_{\rm sim}(i)$ are plotted in Figures 8, 9, and 10 for the
unbroken, sharply-broken, and slowly-bending power-law models, respectively.

In order for the distribution of best-fit parameters to the synthetic data sets to
map out the true parameter probability distributions,
the following conditions must hold:
First, the distribution of $a_{\rm sim}(i) - a_0(i)$
is assumed to be the same as the distribution of $a_{\rm sim}(i) - a_{\rm true}(i)$.
The assumption is reasonable provided that,
over the confidence region, the shape of the distribution does not change significantly.
This likely applies to the current situation:
the assumed values for $f_{\rm b}$ are not close to the edges of the temporal frequency range of the PSD.
In addition, we are not assuming power-law slopes steeper than --1.8;
for power-law slopes steeper than $\sim$--2.5 to --3.0, the large amount of 
red noise leak increases the scatter in the amplitudes of simulated PSDs, thus potentially affecting 
the shape of the distribution as one considers relatively steeper PSD slopes.
Second, the estimator bias, defined as the difference between $a_{\rm 0}(i)$
and the average of the distribution of $a_{\rm sim}(i)$, must be close to $a_{\rm 0}(i)$.
For all power-law slopes, the estimator biases ranged from 0.0 to a maximum of 0.1;
estimator biases for $f_{\rm b}$ were +0.1 in the log for the sharply-broken power-law
and +0.2 for the slowly-bending power-law.
In addition, the distribution of $a_{\rm sim}(i) - a_{\rm 0}(i)$ must be symmetric;
for non-symmetric distributions, a higher probability of obtaining a value
higher or lower than the average of $a_{\rm sim}(i) - a_{\rm 0}(i)$
should be reflected in the reported errors.
Confidence ranges reported in Tables 2 and 3 thus take into account
estimator bias and non-symmetric distributions.


For the unbroken power-law model,
the central 68$\%$ of the distribution, i.e., the 68$\%$ confidence range, spanned 1.30--1.36.
The 99$\%$ confidence ranges span 1.22--1.56, roughly similar to the
error derived by the M03 method. We conclude that the M03 error listed in Table 2 
can be regarded as a roughly 99$\%$ confidence range.
We did not run the P92 method for the 2--5, 5--10, and 10--20 keV
PSD as that would have required an additional 300$\%$ computing time.
However, given that the M03 errors listed in Table 2 are roughly similar to 
that for the 2--10 keV PSD, it is reasonable to assume 
that the error on $\alpha$ listed in Table 2 for each of these PSDs is 
also a roughly 99$\%$ confidence range; for readers interested in quoting 68 or 90$\%$ confidence
ranges, only the P92 ranges are useful.
Derived 68, 90 and 99$\%$ confidence ranges for the parameters of the
broken power-law models are listed in Table 3. Comparing 
these ranges with the M03 error ranges
for the parameters of the sharply-broken power-law model,
we see that the latter tend to roughly the same size as the P92 68--90$\%$ confidence ranges. 

Future PSD work can incorporate errors on model parameters that will be estimated using a method that
utilizes the Neyman construction, whose application to PSD measurement is
currently under development: the reader is referred to Mueller \& Madejski (2009) for details.

\begin{deluxetable}{llcccccc}
\tablecolumns{8}
\tabletypesize{\footnotesize}
\tablewidth{6in}
\tablecaption{Light Curve Sampling and PSD Measurement Parameters \label{tab1}}
\tablehead{
\colhead{}         & \colhead{}     & \colhead{Source flux}             & \colhead{Mean}        & \colhead{Mean}          & \colhead{} \\
\colhead{Bandpass} & \colhead{Time} & \colhead{($10^{-11}$}              & \colhead{Source}      & \colhead{Bkgd}          & \colhead{$F_{\rm var}$}  & \colhead{Temporal Frequency}      & \colhead{$P_{\rm Psn}$} \\
\colhead{(keV)}    & \colhead{Scale} & \colhead{erg cm$^{-2}$ s$^{-1}$)} & \colhead{ct s$^{-1}$}  & \colhead{(ct s$^{-1}$)}  & \colhead{($\%$)}     & \colhead{Range Spanned (Hz)}  & \colhead{(Hz$^{-1}$)} }
\startdata
2--10  & Long   & 3.06 &  2.94  & 3.44  & $22.0 \pm 0.16$  & $8.7 \times 10^{-9} - 1.1 \times 10^{-6}$ & 545   \\
       & Medium & 2.87 &  2.67  & 3.59  & $15.93 \pm 0.12$ &  $6.3 \times 10^{-7} - 7.2 \times 10^{-5}$ & 0.59  \\
       & Short1 &      &  2.463 & 0.015 &  $8.39 \pm 0.22$  & $2.0 \times 10^{-5} - 2.0 \times 10^{-4}$ & 0.98 \\
       & Short2 &      &  2.516 & 0.010 &  $6.06 \pm 0.22$  & $2.2 \times 10^{-5} - 2.0 \times 10^{-4}$ & 0.88 \\
2--5   & Long   & 1.70 &  1.50  & 1.47  & $24.5 \pm 0.22$ & $8.7 \times 10^{-9} - 1.1 \times 10^{-6}$ & 974  \\
       & Medium & 1.56 &  1.37  & 1.62  & $17.20 \pm 0.13$ & $6.3 \times 10^{-7} - 7.2 \times 10^{-5}$ & 1.07 \\
       & Short1 &      &  1.880 & 0.008 & $7.88 \pm 0.24$  & $2.0 \times 10^{-5} - 2.0 \times 10^{-4}$ & 1.24 \\
       & Short2 &      &  1.904 & 0.006 & $6.02 \pm 0.27$   & $2.2 \times 10^{-5} - 2.0 \times 10^{-4}$ & 1.17 \\
5--10  & Long   & 1.53 &  1.52  & 2.04  & $20.5 \pm 0.23 $  & $8.7 \times 10^{-9} - 1.1 \times 10^{-6}$ & 1130 \\
       & Medium & 1.28 &  1.27  & 1.89  & $15.08 \pm 0.13$ &  $6.3 \times 10^{-7} - 7.2 \times 10^{-5}$ & 1.31  \\
       & Short1 &      &  0.584 & 0.007 & $9.98 \pm 0.46$  & $2.0 \times 10^{-5} - 2.0 \times 10^{-4}$ & 3.37 \\
       & Short2 &      &  0.613 & 0.004 & $6.20 \pm 0.47$  & $2.2 \times 10^{-5} - 2.0 \times 10^{-4}$ & 4.09 \\
10--20 & Long   & 1.72 &  0.50  & 2.14  & $19.92 \pm 0.64$  & $8.7 \times 10^{-9} - 1.1 \times 10^{-6}$ & 7770  \\
       & Medium & 1.70 &  0.46  & 2.38  & $13.61 \pm 0.40 $ & $6.3 \times 10^{-7} - 7.2 \times 10^{-5}$ & 9.06   
\enddata
\tablecomments{Source fluxes are corrected for Galactic absorption.
\textit{XMM-Newton} fluxes were estimated from pn count rates using the HEASARC's online W3PIMMS tool and assuming a power-law
photon index of 1.7. 
Count rates are in units of ct s$^{-1}$ PCU$^{-1}$ for the long- and medium-term \textit{RXTE} sampling
or pn ct s$^{-1}$ for the short-term \textit{XMM-Newton} light curves.
For each long-, medium- and all short-term light curves, the sampling time $\Dtsamp$ was  368.6 ks (4.266 days), 5775 s, and 2000 s, respectively.
For each of the long, medium, short1, and short2 light curves, the durations were 2290 days, 32.0 days,
85.0 ks, and 79.0 ks, respectively.
The temporal frequency ranges spanned refer to the binned PSD, not the unbinned periodogram.
The values of $P_{\rm Psn}$ for the short-term light curves were measured from the high-frequency
PSDs above $10^{-3.6}$ Hz, constructed from light curves binned to 300 s. The values of $P_{\rm Psn}$
for the medium- and long-term light curves were estimated based on the net and background count rates (see the text
for details).}
\end{deluxetable}

\begin{deluxetable}{lcccc}
\tablecolumns{5}
\tabletypesize{\footnotesize}
\tablewidth{6in}
\tablecaption{Unbroken Power Law Model Fits to PSDs\label{tab3}}
\tablehead{
\colhead{Bandpass}     &   \colhead{}                  &   \colhead{}                  &   \colhead{}     &   \colhead{$L_{\rm unbr}$}    \\
\colhead{(keV)}        &   \colhead{$\alpha$}          &   \colhead{log($A_0$,Hz$^{-1}$)} &   \colhead{$\xsqdist$/dof}           &   \colhead{($\%$)}  }
\startdata
2--10                  & $1.36\pm0.20$           &  3.62 $\pm$ 0.04 &   63.5/35 &   1.1    \\
                       & (68$\%$: 1.30--1.36)    &                  &          &             \\
                       & (90$\%$: 1.26--1.40)    &                  &          &             \\
                       & (99$\%$: 1.22--1.56)    &                  &          &             \\
2--5                   & $1.40^{+0.22}_{-0.32}$    &  3.70 $\pm$ 0.04 &   56.2/35 &   1.8    \\
5--10                  & $1.44^{+0.12}_{-0.28}$    &  3.58 $\pm$ 0.04 &   70.6/35 &   1.6    \\
10--20                 & $1.20^{+0.08}_{-0.22}$    &  3.51 $\pm$ 0.04 &   28.1/23 &   9.9      
\enddata
\tablecomments{Results from fitting the PSDs with unbroken power law model.
$A_0$ is the normalization at $1 \times 10^{-6}$ Hz. 
$L_{\rm unbr}$ is the likelihood of acceptance for this model,
defined as one minus the rejection probability.
The uncertainty on $\alpha$ listed next to the best-fit value
for the 2--10 keV PSD is the 1$\sigma$ error estimated using the method described in M03.
The 68, 90 and 99$\%$ confidence ranges listed in parentheses were estimated using the P92 method.}
\end{deluxetable}

\begin{deluxetable}{lcccccc}
\tablecolumns{6}
\tabletypesize{\footnotesize}
\tablewidth{6in}
\tablecaption{Broken Power Law Model Fits to PSDs\label{tab4}}
\tablehead{
\colhead{Bandpass}  &   \colhead{}                  &  \colhead{$f_{\rm b}$}    & \colhead{}                  & \colhead{}                  & \colhead{}  & \colhead{$L_{\rm brkn}$}   \\    
\colhead{(keV)}     &   \colhead{$\alpha_{\rm lo}$}  &  \colhead{($10^{-6}$ Hz)} & \colhead{$\alpha_{\rm hi}$}  & \colhead{log($A_1$,Hz$^{-1}$)} & \colhead{$\xsqdist$/dof}          & \colhead{($\%$)}   }
\startdata
\multicolumn{6}{c}{Sharply Broken Power-Law} \\   \hline
2--10   & $0.8\pm0.3$       &  $2.0^{+2.0}_{-1.2}$   &  $1.8 \pm 0.2$      & 3.79$\pm$0.03 &   29.3/33  &  71.8 \\
        & (68$\%$: 0.7--0.9)      & (68$\%$: 1.26--5.01)         & (68$\%$:1.7--2.0)       &                &        &             \\  
        & (90$\%$: 0.6--1.0)      & (90$\%$: 0.79--5.01)         & (90$\%$:1.7--2.1)       &                &        &             \\  
        & (99$\%$: 0.3--1.1)      & (99$\%$: 0.63--7.94)         & (99$\%$:1.6--2.2)       &                &        &             \\  
2--5    & $0.9^{+0.3}_{-0.2}$ & $4.0^{+4.0}_{-2.4}$    &  $1.9^{+0.2}_{-0.3}$ & 3.45$\pm$0.03 &   39.5/33 &  30.2 \\
5--10   & $0.9\pm0.3$       & $4.0^{+6.0}_{-2.4}$    &  $1.9^{+0.2}_{-0.3}$ & 3.33$\pm$0.03 &   32.4/33 &  58.4 \\
10--20  & $1.1^{+0.1}_{-0.2}$ & $2.5^{+10.1*}_{-2.3*}$ &  $1.2^{+0.3}_{-0.2}$ & 3.11$\pm$0.04 &   23.8/21 &  23.4 \\ \hline
\multicolumn{6}{c}{Slowly Bending Power-Law} \\   \hline
2--10   & $0.5^{+0.4}_{-0.2}$ & $1.0 \pm 0.6$       &  $1.8 \pm 0.2$      & 1.38$\pm$0.03 &  27.4/33 &  80.6  \\
        & (68$\%$: 0.2--0.8)      & (68$\%$: 0.40--3.98)         & (68$\%$:1.7--2.0)       &                &        &             \\  
        & (90$\%$: 0.0--0.9)      & (90$\%$: 0.25--7.94)         & (90$\%$:1.6--2.3)       &                &        &             \\  
        & (99$\%$: 0.0--1.0)      & (99$\%$: 0.13--12.6)         & (99$\%$:1.6--2.5)       &                &        &             \\  
2--5    & $0.8^{+0.3}_{-0.2}$ & $10.0^{+5.8}_{-5.0}$   &  $2.3 \pm 0.4$    & --1.31$\pm$0.03  & 33.8/33    &  52.1  \\
5--10   & $0.8^{+0.2}_{-0.3}$ & $5.0^{+5.0}_{-2.5}$   &  $2.1 \pm 0.3$     & --0.79$\pm$0.03  & 27.9/33   &  77.1  \\
10--20   & $1.0^{+0.3}_{-0.4}$ & $0.4^{+4.6}_{-0.2*}$  &  $1.2^{+0.1}_{-0.2}$ & --2.31$\pm$0.04  &   23.1/21 &  24.0    
\enddata
\tablecomments{Results from fitting the PSDs with two broken power law models,
the sharply-broken (top half) and slowly-bending (bottom half) models.
$\alpha_{\rm lo}$ and $\alpha_{\rm hi}$ are the
power law slopes below and above $f_{\rm b}$, respectively.  $A_1$ is the PSD normalization. $L_{\rm brkn}$ 
is the likelihood of acceptance for this model, defined as one minus the rejection probability.
An asterisk (*) denotes that the parameter uncertainty pegged at the limit of the parameter space tested.
The uncertainties on $\alpha_{\rm lo}$, $f_{\rm b}$, and $\alpha_{\rm hi}$ listed next to the best-fit values
for the 2--10 keV PSD are the 1$\sigma$ errors estimated using the M03 method.
The 68, 90 and 99$\%$ confidence ranges listed in parentheses were estimated using the P92 method.}
\end{deluxetable}

\begin{deluxetable}{llcc}\tablecolumns{4}
\tabletypesize{\footnotesize}
\tablewidth{0pc}
\tablecaption{Broadband \textit{RXTE} PCA+HEXTE Spectral Fits\label{tab4}}
\tablehead{
\colhead{}          & \colhead{}          & \colhead{1996/Medium}  &   \colhead{2003--2009/Long} \\             
\colhead{Component} & \colhead{Parameter} & \colhead{Value}   & \colhead{Value}    }
\startdata
             & $\chi^2$/dof                             &   85.0/77 &   66.3/69 \\        
Model flux   & 2--10 keV Flux (erg cm$^{-2}$ s$^{-1}$)   &   $ 3.24 \times 10^{-11}$ &   $ 3.26 \times 10^{-11}$ \\
Model flux   & 10--20 keV Flux  (erg cm$^{-2}$ s$^{-1}$)   &    $ 1.89 \times 10^{-11}$ &   $ 1.89 \times 10^{-11}$ \\
Model flux   & 20--100 keV Flux (erg cm$^{-2}$ s$^{-1}$) &    $ 4.43 \times 10^{-11}$ &   $ 4.25 \times 10^{-11}$ \\
Galactic absorption &  $N_{\rm H, Gal}$ (cm$^{-2}$)        &    $4.45 \times 10^{20}$ (fixed)  &    $4.45 \times 10^{20}$ (fixed) \\
Power law    & $\Gamma$                                 &  $1.89^{+0.03}_{-0.01}$  & $1.95^{+0.05}_{-0.04}$  \\
             & Norm.\ at 1 keV (ph cm$^{-2}$ s$^{-1}$ keV $^{-1}$) &  $1.00 \pm 0.04 \times 10^{-2}$ &  $1.09 \pm 0.06 \times 10^{-2}$  \\
Compton reflection          & Refl.\ strength $R$                     &  0.48$\pm$0.12                 & $0.68^{+0.21}_{-0.17}$ \\
                            & Reflected/total flux, 10--20 keV        &  0.21                            &  0.27     \\
Fe K$\alpha$ emission line  & Energy (keV)                           & $6.37^{+0.09}_{-0.11}$            &  6.38$\pm$0.09 \\
                            & Width $\sigma$ (keV)                    &  $<$0.48                      & $<$0.41 \\
                            & Intensity (ph cm$^{-2}$ s$^{-1}$)       &   $5.2^{+0.7}_{-0.9} \times 10^{-5}$ &    $5.2^{+1.5}_{-0.8} \times 10^{-5}$ \\
                            & EW (eV)                              &   $150^{+21}_{-27}$  &   $146^{+44}_{-23}$ \\
Good exposure time          & PCA/HEXTE-A/HEXTE-B  (ks)              &  689.4/193.5/196.5  & 375.2/49.7/108.5 
\enddata
\tablecomments{Results from joint fits to 3--30 keV PCA and 15--100 keV HEXTE 
using a model consisting of a power-law component, a Compton reflection component ({\sc pexrav}), and
an Fe K$\alpha$ emission line, all modified by the Galactic column. 20--100 keV fluxes are
weighted between HEXTE clusters A and B accounting for good exposure time and the number of active detectors.
For 2003--2009, the good exposure time for HEXTE-A is much less than that for HEXTE-B because HEXTE-A stopped collecting realtime
background data in 2006.}
\end{deluxetable}

\begin{figure}[ht]
\epsscale{0.7}
\plotone{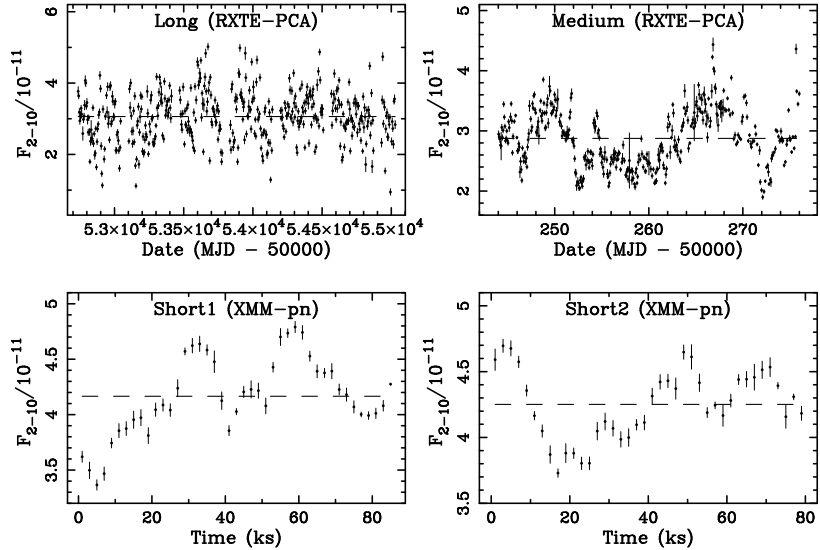}   
\caption{X-ray light curves of NGC 7469 from multi-time scale monitoring.
Plotted are values of 2--10 keV flux in units of 10$^{-11}$ erg cm$^{-2}$ s$^{-1}$.}
\end{figure}

\begin{figure}
\epsscale{0.5}
\plotone{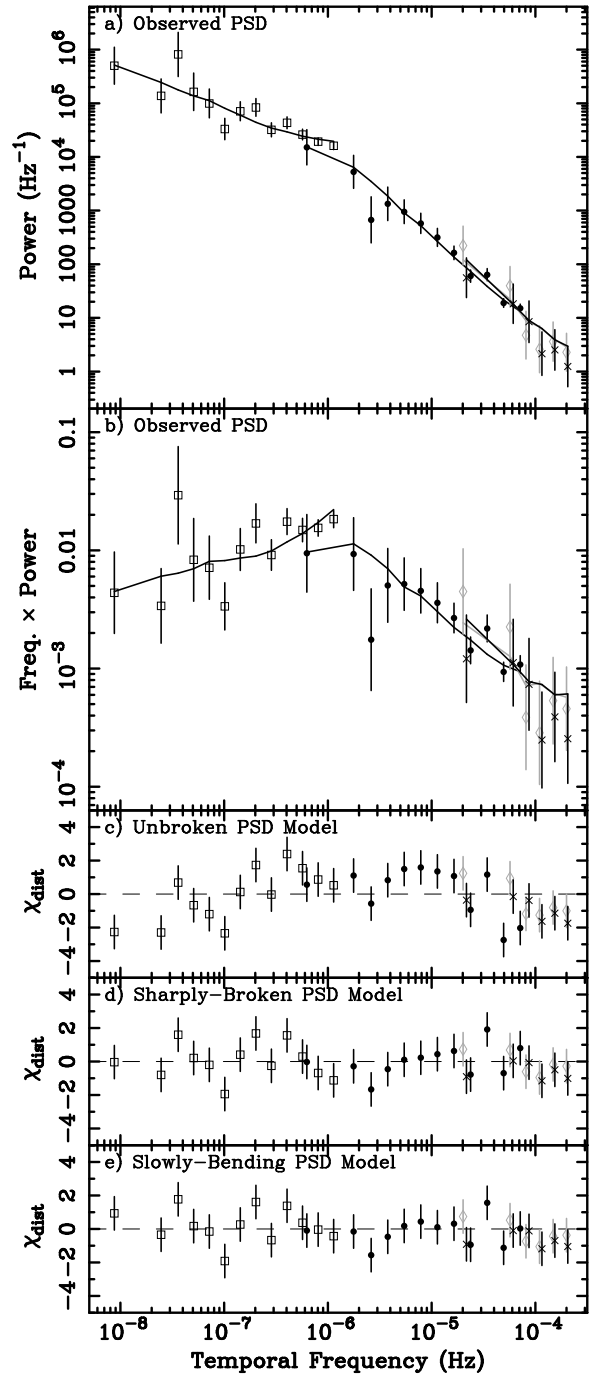}    
\caption{The top two panels show the observed 2--10 keV PSD, plotted in 
$f \times P(f)$ in panel (b) space to visually emphasize the turnover.
Open squares, filled circles, gray open diamonds and crosses
denote the long, medium, short1, and short2 PSD segments, respectively.
The solid lines in panels (a) and (b) denote the best-fit sharply-broken
power-law model folded through the sampling window (i.e., containing the distortion
effects of PSD measurement and power due to Poisson noise).
Panels (c), (d), and (e) show the data--model residuals to the best-fit
unbroken, sharply-broken, and slowly-bending power-law models, respectively.}
\end{figure}

\begin{figure}
\epsscale{0.85}
\plotone{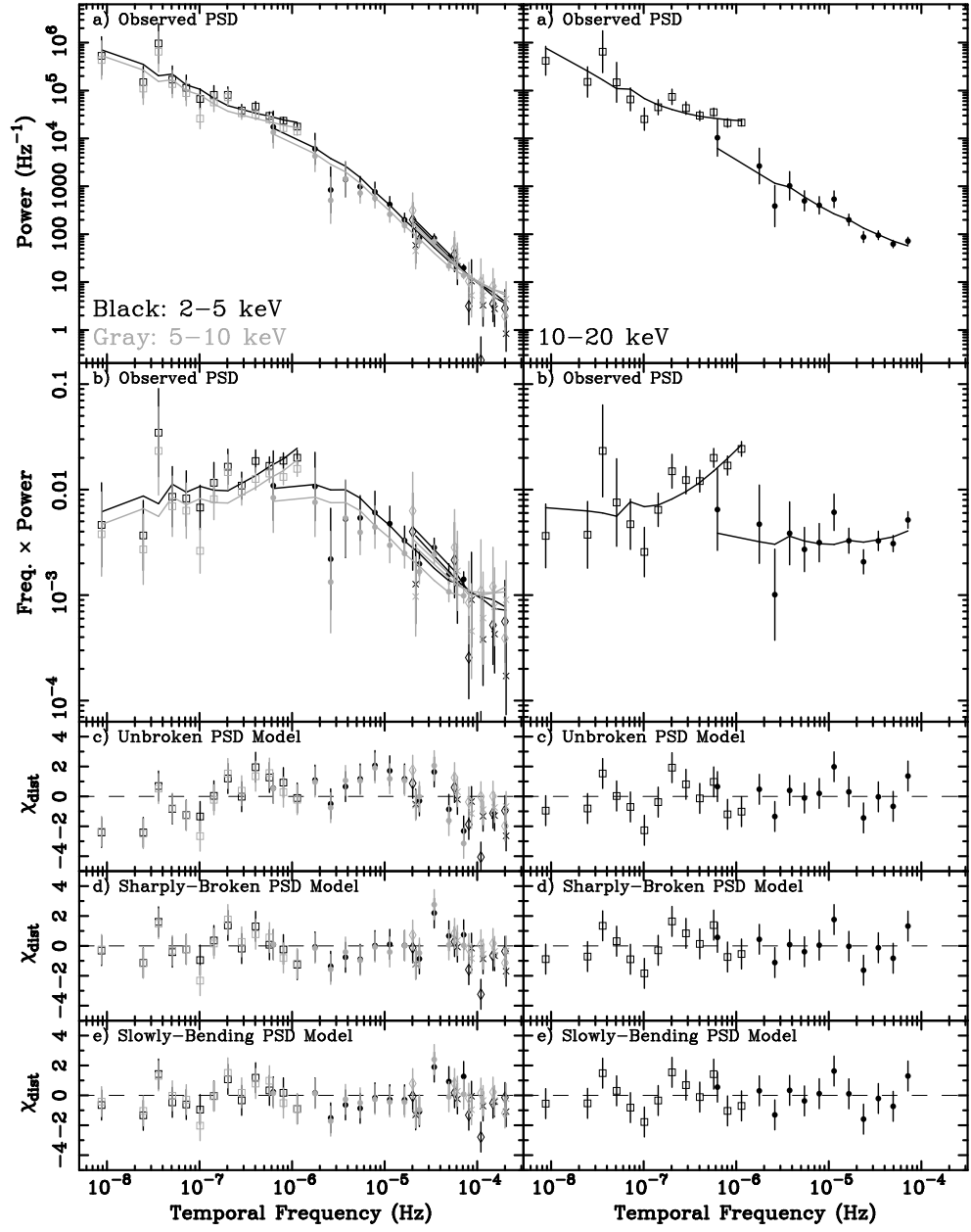}    
\caption{Same as Figure 2, but for the 2--5 (left, black points), 
5--10 (left, gray points) and 10--20 keV sub-bands (right).}
\end{figure}

\begin{figure}
\epsscale{0.85}
\plotone{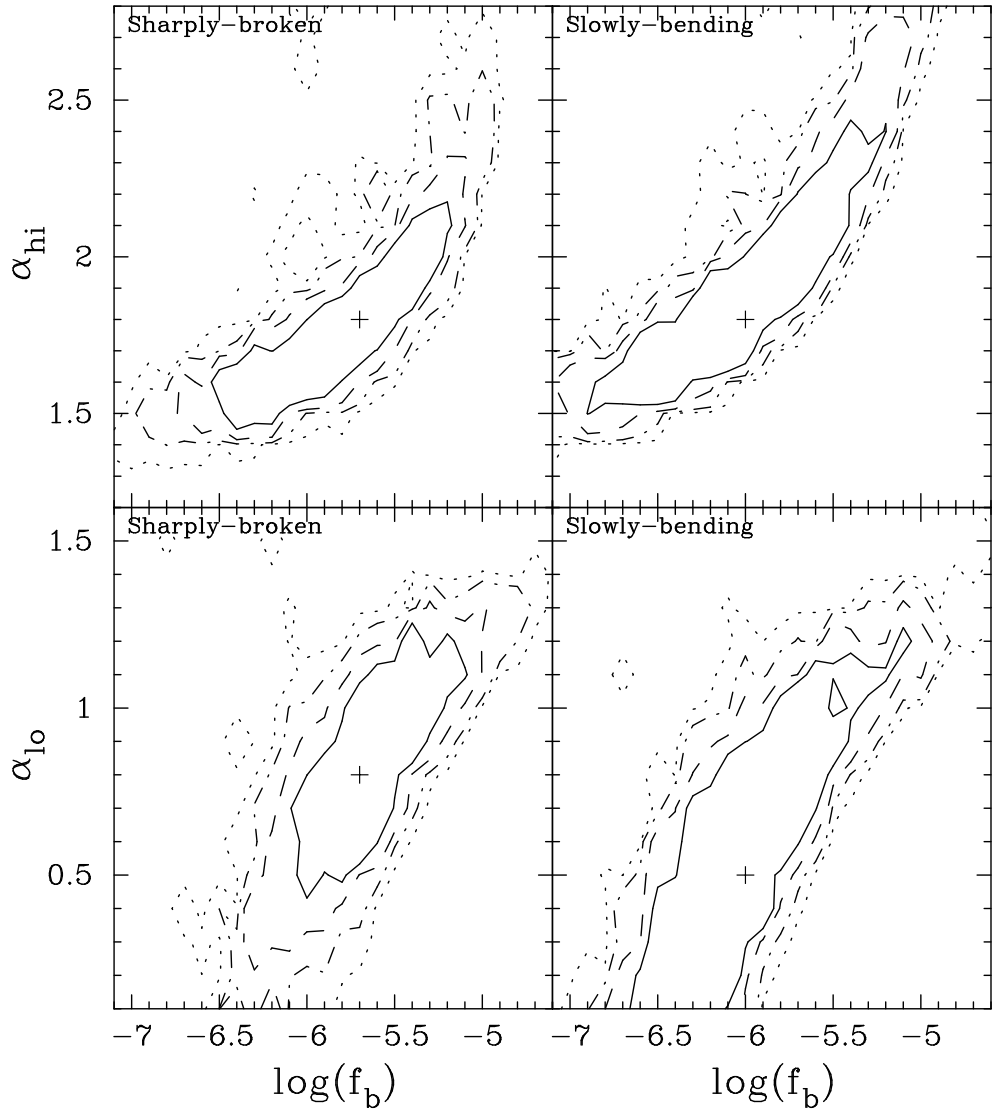}      
\caption{Contour plots of $\alpha_{\rm hi}$ and $\alpha_{\rm lo}$ vs.\ $f_{\rm b}$
for the best-fit sharply-broken (left) and slowly-bending (right) power-law models
for the 2--10 keV PSD.
Each contour represents a slice through the parameter space tested
at the best-fit value of $\alpha_{\rm hi}$ or $\alpha_{\rm lo}$ (listed in Table 3).
Solid, dashed, dot-dashed, and dotted contours denote
68, 90, 95.4, and 99.0$\%$ rejection probabilities, respectively.}
\end{figure}

\begin{figure}
\epsscale{0.85}
\plotone{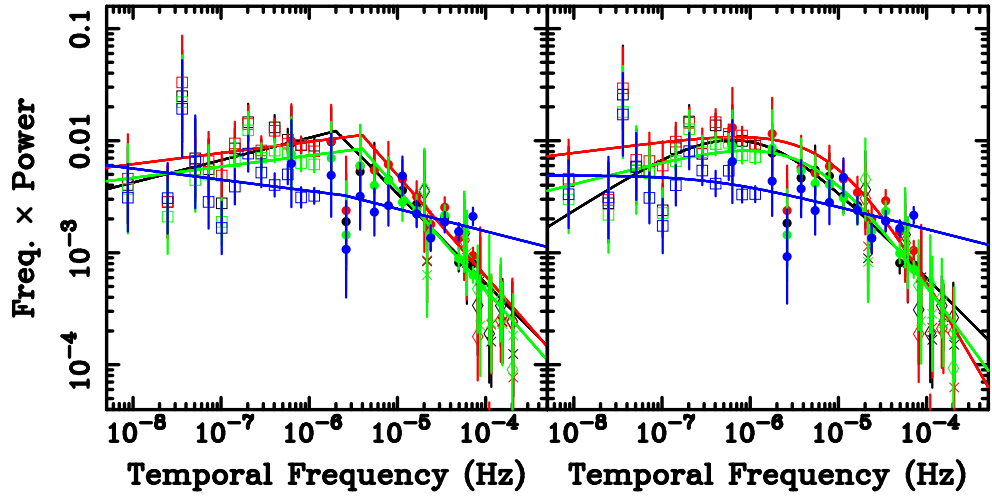}   
\caption{The best-fit sharply-broken (left) and slowly-bending (right) model fits to
the 2--10 (black), 2--5 (red), 5--10 (green), and 10--20 keV (blue) PSDs with the PSD distortion measurement
effects and power due to Poisson noise removed.
The PSDs are plotted in ``model space'': the solid lines denote the best-fit models, and
symbols denote the differences between the observed PSD points and $\overline{P_{{\rm sim}}(f)}$, 
plotted relative to the underlying PSD model.}
\end{figure}

\begin{figure}
\epsscale{0.4}
\plotone{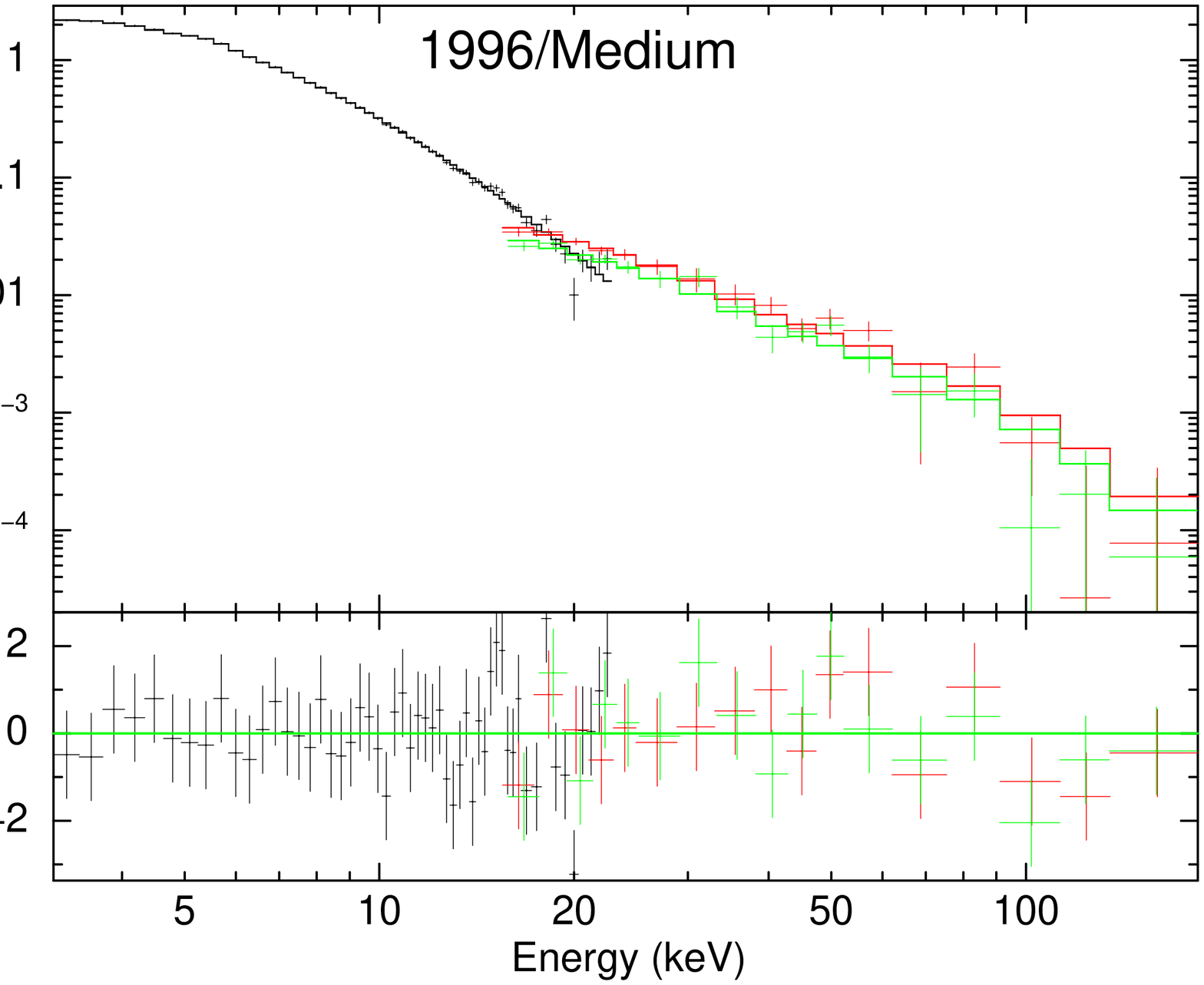}    
\caption{\textit{RXTE} PCA (black points) and HEXTE cluster A and B (red and green points, respectively) time-averaged
spectrum and best-fit model for the 1996/medium-term campaign are shown in the upper panel, and the
data--model residuals are shown in the lower panel.}
\end{figure}

\begin{figure}
\epsscale{0.4}
\plotone{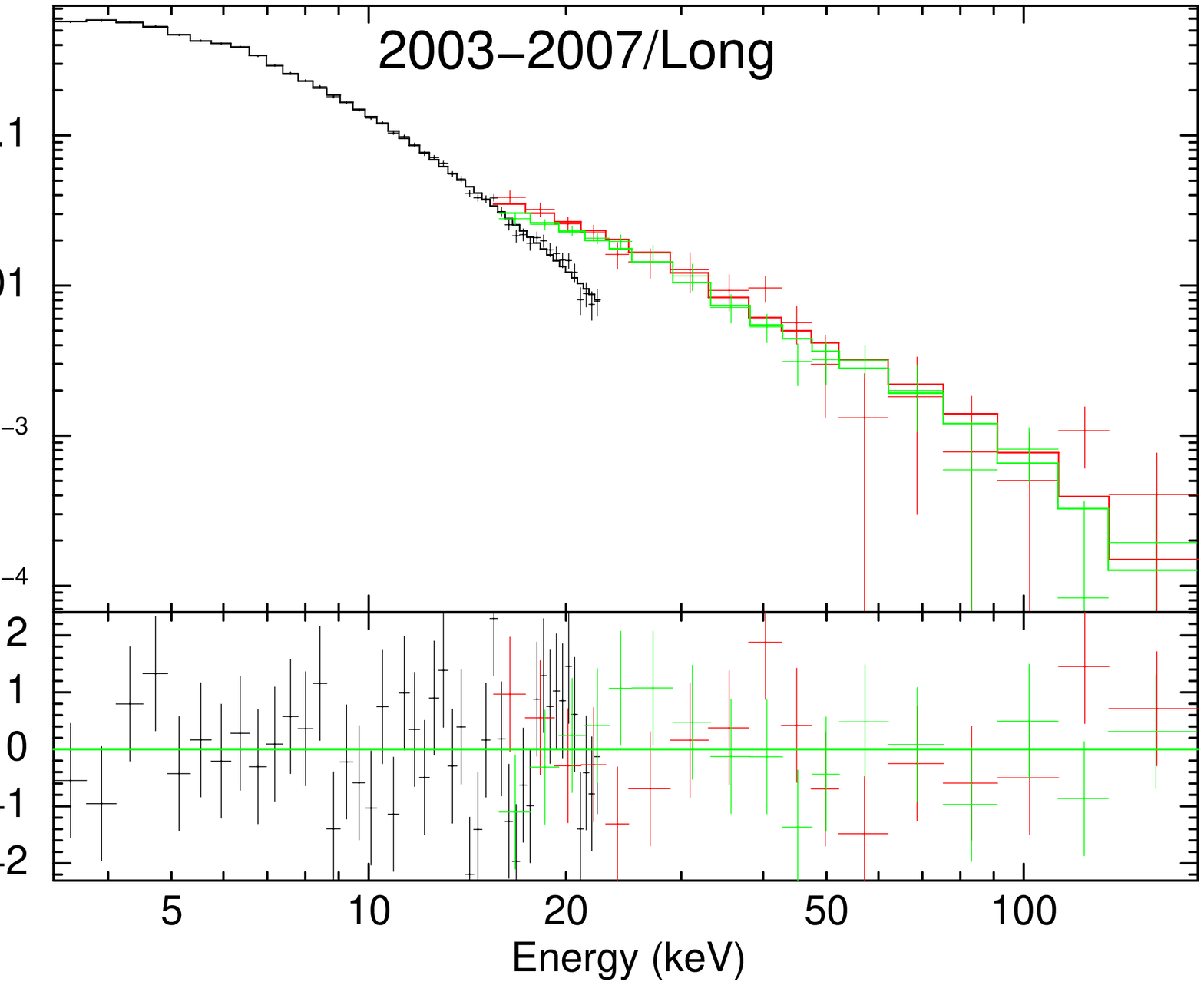}     
\caption{Same as Figure 6, but for the time-averaged 2003--2009/long-term \textit{RXTE} spectrum and best-fit model.}
\end{figure}

\begin{figure}
\epsscale{0.35}
\plotone{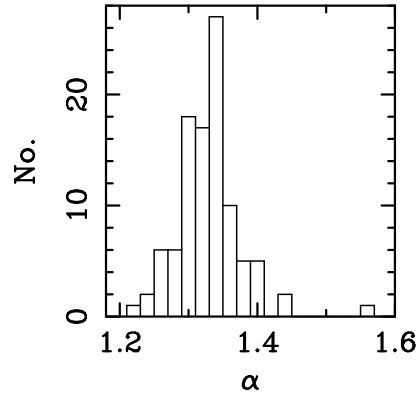}    
\caption{The distribution of best-fit power-law slopes $a_{\rm sim}(i)$ to 100 simulated data sets 
for the best unbroken power-law model fit to the 2--10 keV PSD.
Data sets were simulated assuming a power-law slope $a_{\rm 0}(i)$ of 1.36. 
As the average of the distribution of $a_{\rm sim}(i)$ was 1.32,
the estimator bias is $\langle a_{\rm sim}(i) \rangle - a_{\rm 0}(i)$ = 0.04. The 68, 90 and 99$\%$
confidence regions are reported in Table 2.}
\end{figure}

\begin{figure}
\epsscale{0.7}
\plotone{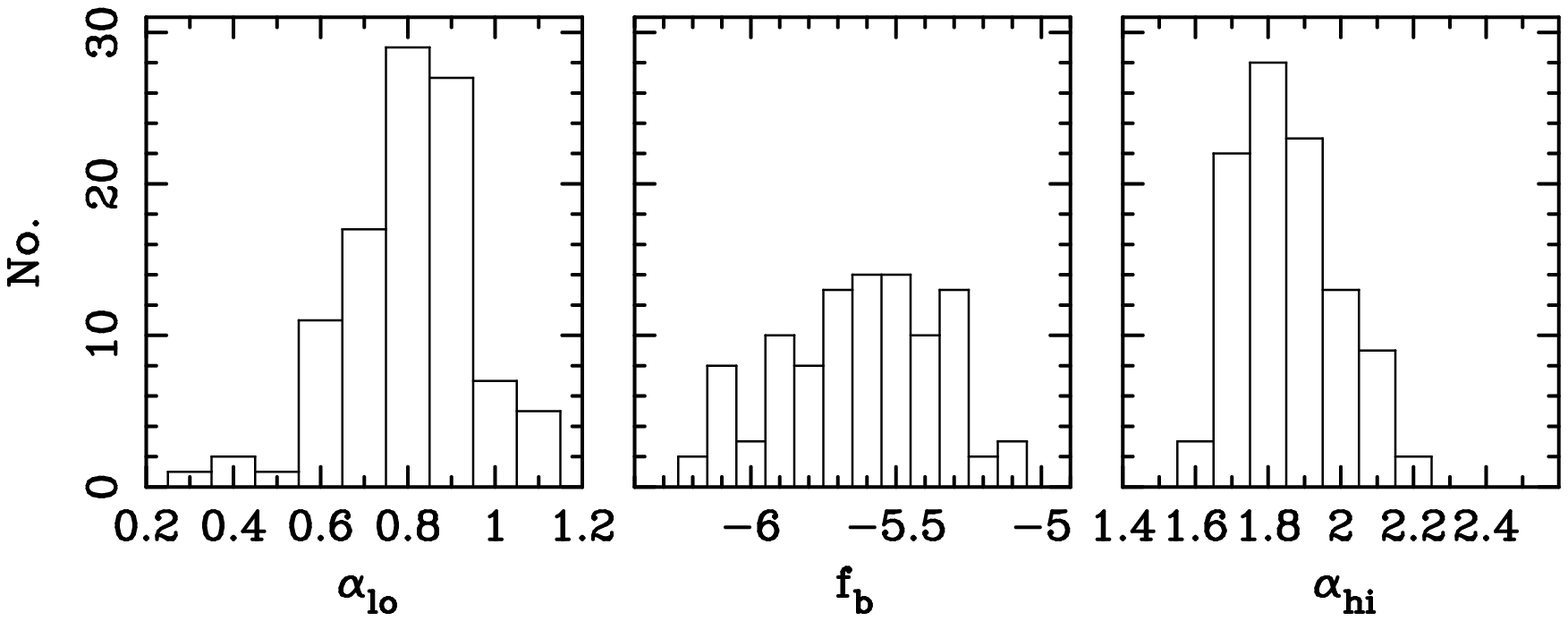}  
\caption{The distribution of best-fit parameters $a_{\rm sim}(i)$ to 100 simulated data sets 
for the best sharply-broken power-law fit to the 2--10 keV PSD.
Data sets were simulated assuming $\alpha_{\rm lo} = 0.8$, $f_{\rm b} =  2 \times 10^{-6}$ Hz, and
$\alpha_{\rm hi} = 1.8$; the resulting estimator biases $\langle a_{\rm sim}(i) \rangle - a_{\rm 0}(i)$ were 
0.0, +0.1 in the log, and +0.1, respectively. The 68, 90 and 99$\%$
confidence regions are reported in Table 3.}
\end{figure}

\begin{figure}
\epsscale{0.7}
\plotone{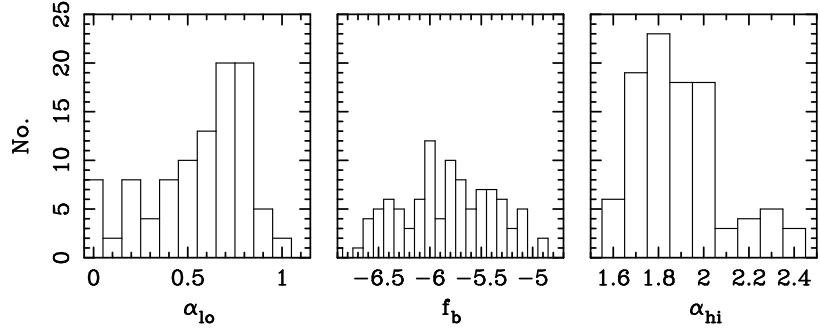}    
\caption{The distribution of best-fit parameters $a_{\rm sim}(i)$ to 100 simulated data sets 
for the best slowly-bending power-law fit to the 2--10 keV PSD.
Data sets were simulated assuming $\alpha_{\rm lo} = 0.5$, $f_{\rm b} =  1 \times 10^{-6}$ Hz, and
$\alpha_{\rm hi} = 1.8$; the resulting estimator biases $\langle a_{\rm sim}(i) \rangle - a_{\rm 0}(i)$ were 
+0.1, +0.2 in the log, and +0.1, respectively. The 68, 90 and 99$\%$
confidence regions are reported in Table 3.}
\end{figure}

\end{document}